**Magnetic hydrogels derived from polysaccharides with improved specific power absorption: potential devices for remotely triggered drug delivery.**

R. Hernández[1,*], J. Sacristán[1], L. Asín[2], T. E. Torres[2,3], M. R. Ibarra[2,3] G. F. Goya[2,3], and C. Mijangos[1]

[1]*Instituto de Ciencia y Tecnología de Polímeros, CSIC, Juan Cierva 3, E-28006 Madrid, Spain.*

[2] *Instituto de Nanociencia de Aragón (INA), Universidad de Zaragoza, 50018, Spain.*

[3] *Departamento de Física de la Materia Condensada, Universidad de Zaragoza, Zaragoza, Spain.*

**Abstract**

We report on novel ferrogels derived from polysaccharides (sodium alginate and chitosan) with embedded iron oxide nanoparticles synthesized in situ and their combination with thermally responsive poly (N-isopropylacrylamide) for externally-driven drug release using AC magnetic fields. Samples were characterized by Raman spectroscopy, transmission electron microscopy (TEM) and magnetic measurements. The obtained nanoparticles were found to be of ca. 10 nm average size, showing magnetic properties very close to those of the bulk material. The thermal response was measured by power absorption experiments, finding specific power absorption (SPA) values between 100-300 W/g, which was enough for attaining the lower critical solution temperature (LCST) of the polymeric matrix within few minutes. This fast response makes these materials good candidates for externally controlled drug release.

## 1. Introduction

The development of new controlled release formulation techniques has been boosted by the discovery of new biocompatible polymers and nanomaterials. The bottom line is, regarding clinical applications, to enhance patient compliance and to produce improved drug delivery profiles. Controlled release systems based on polymer gels and microgels have opened the possibility of remote triggering controlled by magnetic fields by incorporating magnetic nanoparticles (NPs), and thus making possible non-invasive release of therapeutic molecules at specific points deeply located in the body[1-3].

The relatively new technique known as magnetic hyperthermia aims the raising of the temperature of a magnetic colloid (or targeted cells) from the magnetic coupling between the magnetic moment of the NPs and the applied alternate magnetic field[2]. The ability of generating heat is measured by the specific power absorption (SPA), which is the power absorbed per unit mass of magnetic nanoparticles. Regarding the physical mechanisms behind heat generation by power absorption, some details of the model proposed by Rosensweig [4] are still being discussed [5,6]. The heating agents (i.e., the magnetic nanoparticles) have been also used to rise the temperature of different target systems, such as polymers [7] and membranes [1,8,9] and, as the largest field of application, cancer cells [10,11].

The combination of magnetic hyperthermia and controlled drug release is currently under intensive investigation for cancer treatments[12]. By combining magnetic nanoparticles with thermosensitive polymers, an alternating current (AC) magnetic field can be used to trigger localized heating *in vivo*, which in turn causes a phase change in the host polymer to allow diffusion and release of drugs [13-15]. Among the temperature responsive polymers, one of the most popular for controlled drug release applications is the poly(*N*-isopropylacrylamide) (PNIPAAm) as it exhibits a lower critical solution temperature (LCST) that can be tailored to be within the physiological range, i.e., human body temperature[16, 17]. The application of an alternating magnetic field on PNIPAAm hydrogels loaded with $Fe_3O_4$ nanoparticles (20-30 nm) results in uniform heating of the ferrogel leading to accelerated collapse and squeezing out large amounts of imbibed drug (release at a faster rate). In addition, the drug dose can be modulated with the application time of the alternating magnetic field[18].

We have recently described a method to obtain ferrogels with response to temperature and magnetic fields, in which the iron oxide NPs synthesis is carried out through coprecipitation of iron salts in alkaline solutions inside semiinterpenetrating (semiIPN) polymer networks constituted of alginate and PNIPAAm [19]. These ferrogels exhibit an improved deswelling rate with respect to pure PNIPAAm. In addition, the synthesis *in situ* of iron oxide NPs inside Alg-PNIPAAm semi-IPNs allows controlling the polydispersity of the iron oxide NPs when compared to the reaction carried out in an alginate solution. In this way, the polymeric gel acts as a spatial framework to control the iron oxide NPs size distribution[20]. This method has also been employed for the *in situ* synthesis of iron oxide NPs inside chitosan ferrogels obtaining systems with response to pH and magnetic fields[21].

In this work, we present results on magnetic properties and heating experiments on hydrogels derived from polysaccharides (chitosan and alginate), aimed to optimize the physical absorption of magnetic power and subsequent heat release by varying the polymer composition and the nanoparticles content. The composite ferrogels have shown good response to external time-varying magnetic fields, opening the possibility of using them as smart drug delivery materials for bioapplications.

## 2. Experimental part

### 2.1. Preparation of ferrogels

Semi-interpenetrating polymer networks constituted by alginate and PNIPAAm were obtained as reported elsewhere [19, 20]. The synthesis of iron oxide nanoparticles inside Alg-PNIPAAm semi-IPNs hydrogels involves two steps. As a first step, gels are immersed in an acid solution containing $Fe^{2+}$ and $Fe^{3+}$. As a second step, the solutions are carefully immersed in an alkaline solution to oxidize the iron cations to iron oxide nanoparticles. For this work, two series with the same degree of crosslinking of the PNIPAAm ($mol_{BIS}/mol_{N-AAm}$= 3 %) and different alginate concentrations were synthesized, **series I** with 0.5 wt% sodium alginate (Alg0.5-PNIPAAm) and **series II** with 1 wt% sodium alginate (Alg1-PNIPAAm). For each of these series, the oxidation procedure was repeated up to three times. The resulting gels were designated as Alg-PNIPAAm (1c) and Alg-PNIPAAm (3c) where the numbers in brackets denote the number of oxidation cycles (see Table I).

Representative TEM images taken on dried ferrogels with different alginate content (0.5 and 1 wt%) subjected to one oxidation cycle indicate that the iron oxide NPs size is independent from the alginate concentration being around 10 nm for both samples **(figure 1)**. A recent study indicated that the particle size slightly increased with the application of repeating oxidation cycles[20].

A similar procedure was carried out for the preparation of chitosan ferrogels (**Series III**)[21]. Briefly, an aqueous solution of iron ions (0.21g of $FeCl_2 \cdot 4H_2O$ and 0.58 g of $FeCl_3 \cdot 6H_2O$) was added under vigorous stirring to a chitosan solution (4 wt%) in a $N_2$ atmosphere to obtain a final chitosan concentration of 2 wt %. The solution was cast on a teflon mould and carefully immersed in a 1 M sodium hydroxide solution in a $N_2$ atmosphere. The immersion time in NaOH was 30 minutes and 60 minutes, and the resulting gels were designated as Chit2_30min and Chit2_60min, respectively. TEM images reported elsewhere[21] showed a good dispersion of spherical iron oxide nanoparticles with an average size of 10 nm for 2 wt% chitosan ferrogels.

The mass of iron oxide nanoparticles in the ferrogels ($W_{NPs}$) was obtained from thermogravimetric analysis of dried ferrogels [19, 21]. The total water content, $W_t$, was calculated from the weight of the hydrated gels ($M_s$) and the weight of the dried gels ($M_{dry}$) as:

$$W_t = M_s - M_{dry} / M_{dry} - M_{NPs} \qquad (1)$$

The data corresponding to $W_{NPs}$ and $W_t$ are reported in table I.

### *2.2. Magnetic characterization*

Magnetic measurements as a function of temperature and field of the ferrogels were performed in a commercial SQUID magnetometer (MPMS-XL Quantum Design). Zero-field-cooled (ZFC) and field-cooled (FC) curves were measured between 5 K and 300 K. Data were obtained by first cooling the sample from room temperature in zero applied field (ZFC process) to the base temperature. Then a magnetic field of $H_{FC} = 100$ Oe was then applied and the magnetization was measured with increasing temperature up to $T = 300$ K. After the last point was measured, the sample was cooled again to the base temperature keeping the same field (FC process). Magnetization versus temperature data were then measured over increasing temperatures. Hysteresis loops were measured at 5 K and 300 K for applied fields $-5\,T < H < 5\,T$.

### *2.3. Remote heating on an alternating magnetic field*

Specific power absorption (SPA) measurements were done using a home made ac applicator working at 260 kHz and field amplitudes from up to 16 mT, and equipped with an adiabatic sample space (~0.5 ml) for measurements in liquid phase.

Temperature data was taken using a fiber optic temperature probe (Reflex™, Neoptix) immune to rf environments.

*2.4. Raman spectroscopy*

Raman spectra were collected using a Renishaw InVia reflex Raman microscope. The Raman scattering was excited with a 785 nm near-infrared diode laser. A 100×, NA90 objective lens was used, giving a laser spot diameter of ~1 μm and the sampling depth is estimated to be around 1.5 μm (half-width of the confocal depth profile for a silicon wafer). In order to avoid thermal decomposition/degradation of the samples, laser intensity at the sample was reduced to 0.5-1%. Spectra were obtained for a 20-s exposure of the CCD detector in the region 100–1800 $cm^{-1}$ using the extended scanning mode of the instrument and the number of data acquisition cycles that make up the measurement was 2. Raman point maps were acquired using a 100× objective, collecting spectra with 10-20 s exposure time and subsequently moving the sample in a raster pattern with a step size of at least 10 μm. The laser intensity at the sample was reduced to 0.5% to avoid burning the sample. Data acquisition covered the spectral range 1800–100 $cm^{-1}$ with a spectral resolution of 4 $cm^{-1}$ with each exposure of the CCD detector. Raman spectra were recorded and subsequently analysed by using Wire software (Renishaw).

**RESULTS AND DISCUSSION**

1. **Molecular structure of iron oxide NPs**

As it is well known, wide angle X ray diffractograms corresponding to magnetite and maghemite are almost identical and therefore this technique does not allow a detailed characterization of the molecular structure of iron oxide NPs synthesized inside semiIPN of alginate and PNiPAAm[22]. That is the reason why we employed Confocal

Raman spectroscopy to determine the molecular structure of iron oxide NPs which will have a great influence on the magnetization properties of the composite ferrogel as we will show later on.

Representative spectra corresponding to series II (1 wt% alginate) are shown in **Figure 2.** The Raman spectrum corresponding to the sample without iron oxide NPs (Alg1-PNiPAAm) is also shown for comparison. The low frequency region (300-800 cm$^{-1}$) contains the most characteristic Raman bands of iron oxides. The presence of iron oxide NPs in the samples subjected to 1 and 3 oxidation cycles, Alg1-PNIPAAm (1c) and Alg1-PNIPAAm (3c) respectively, is revealed by the appearance of two new peaks at 500 cm$^{-1}$ and 700 cm$^{-1}$. The most intense band located at 700 cm$^{-1}$ can be deconvoluted into two peaks centered at 670 cm$^{-1}$ and 720 cm$^{-1}$ assigned to magnetite and maghemite respectively. The bands detected at 500 and 670 cm$^{-1}$ can be assigned to maghemite, $\gamma$-$Fe_2O_3$ whereas the presence of the third band at 720 cm$^{-1}$ is attributed to magnetite [23, 24].

For the sample subjected to 3 oxidation cycles, Alg1-PNiPAAm (3c), the most representative Raman bands of iron oxide NPs, underwent a nearly four-fold increase in intensity with respect to the sample subjected to 1 oxidation cycle, Alg1-PNIPAAm (1c). The significant presence, in all the samples analyzed, of the 720 cm$^{-1}$ and 670 cm$^{-1}$ bands is indicative of the simultaneous formation of magnetite and maghemite from the oxidation step of the nanoparticle fabrication. These nanoparticles should be considered as composite magnetite and maghemite systems. The ratio of the intensity of the 720/670 cm$^{-1}$ bands provide a semi-quantitative estimation of the maghemite/magnetite content in the sample under study. The ratio $I_{720}/I_{670}$ changes from 1.1 for samples subjected to one oxidation cycle to 1.3 for samples subjected to 3 oxidation cycles. Although not shown here, a similar behaviour is observed in series I, where $I_{720}/I_{670}$ changes from 1.7 to 2.1 for samples subjected to one and three oxidation cycles respectively. This suggests an increase in the maghemite content of nanoparticles as a function of the oxidizing cycles. Presumably the oxidation takes place at the surface of the nanoparticle progressing towards the center and originating, in this case, NPs with large magnetocrystalline anisotropy[25, 26].

The Raman spectra of series III (chitosan) are shown in **Figure 3.** The spectrum corresponding to the sample without iron oxide NPs is also shown for comparison being comprised of broad bands, involving mainly ν(C-O) and ν(C-C) stretching. In this case the presence of iron oxide nanoparticles is assessed by the appearance of one peak at 310 $cm^{-1}$, and a broad band at 705 $cm^{-1}$. Following the same procedure as in the case of series I and II the band detected at 705 $cm^{-1}$ can be deconvoluted into two peaks centered at ~670 $cm^{-1}$ and ~720 $cm^{-1}$ which are assigned to magnetite and maghemite respectively. In addition to these bands, the presence of a third band at 310 $cm^{-1}$ should be attributable to maghemite. However, in this series the band observed in figure 3 at 500 $cm^{-1}$, associated to maghemite, it is not observed because is totally overlapped with some characteristic bands and shoulders from pure chitosan.

In series III the ratio $I_{720}/I_{670}$ changes from 0.9 for samples subjected to one oxidation cycle to 1.1 for samples subjected to 3 oxidation cycles, which clearly indicates that the concentration of magnetite with respect to maghemite maintains almost constant with the immersion time in alkaline solution.

These results demonstrate a strong influence of the method of preparation on the maghemite/magnetite ratio. The application of a repeating number of oxidation cycles in Alg-PNiPAAm semiIPNs (series I and II) gives rise to the increase of the maghemite content. However, in chitosan ferrogels (series III) the maghemite/magnetite ratio does not get affected by the increase in the immersion time in an alkaline solution. These results have an important influence on the heating of these samples when subjected to alternating magnetic fields as we will demonstrate later on.

**Magnetic properties**

The magnetic parameters of the magnetic ferrogels were evaluated in relation to their efficiency as heating elements. As a general feature, magnetization measurements as a function of temperature (ZFC-FC) and hysteresis loops between T = 5 K and 300 K for all samples showed a single-domain behavior of the constituent NPs, having blocking temperatures at about $T_B$=100 K. Representative hysteresis curves for samples Alg0.5-PNiPAAm (1c) and Alg1-PNiPAAm (1c) are shown in (Figure 4). The coercitivity values extracted from M(H) curves (see Table II) were found to be essentially zero at room temperature, in agreement with the blocking temperatures of $T_B \approx$ 70-160 K found in all samples. Therefore, at room temperature, the NPs that are inside the ferrogels are in the SPM state. At low temperatures, both the saturation magnetization $M_S$ and the coercivity $H_C$ values obtained were comparable to previously reported data on similar magnetite (or maghemite) nanoparticles.[27] Because of the similar size (and thus similar surface/core ratio) of the NPs in all ferrogels, the contribution of surface disorder to the magnetic moment of the NPs are expected to be similar.

Magnetization curves as a function of temperature taken in zero-field-cooling and field-cooling modes for all samples confirmed the SPM behavior at room temperature with blocking temperatures ranging from 60 to 160 K (see Table II). Representative ZFC-FC curves for samples Alg0.5-PNiPAAm (1c) and Alg1-PNiPAAm (1c) are shown in (Figure 5).

The observed differences in the values of $T_B$ obtained from M (T) data can be attributed to differences in the aggregation state in each sample, since the aggregation of NPs is known to modify the contribution from dipolar interactions to the anisotropy energy barrier, thus modifying the relaxation times of the nanoparticles [27, 28].

## 3. Remote heating on the application of an alternating magnetic field

We studied the capability of the synthesized ferrogels to remotely trigger the release process by evaluating their heating performance when submitted to an alternating magnetic field. Specifically, we measured the SPA of the NPs synthesized inside the ferrogels through temperature vs time profiles. The heating efficiency of the colloids was determined from the temperature increase ($\Delta T$) of a given mass of the constituent nanoparticles ($m_{NP}$) diluted in a mass of liquid carrier ($m_{LIQ}$) during the time interval ($\Delta t$) of the experiment. The expression for power absorption $P$ per unit mass of the magnetic material is given by:

$$\Pi = \frac{P}{m_{NP}} = \frac{m_{LIQ} c_{LIQ} + m_{NP} c_{NP}}{m_{NP}} \left( \frac{\Delta T}{\Delta t} \right) \qquad (3)$$

where $c_{LIQ}$ and $c_{NP}$ are the specific heat capacities of the liquid carrier and the nanoparticles, respectively. All experiments were done after a waiting time of 60-120 s with the magnetic field off, to allow thermalization of the samples to room temperature. Since the time dependence of temperature T is not linear, the slope of the T(t) is also a function of time. This is usually due to heat losses of the experimental setup, and thus a criterion is needed in order to extract reproducible information from experiments. We have chosen the criterion of the maximum derivative for calculating our SPA values, since this criterion has two main advantages: first, the maximum slope $\Delta T/\Delta t$ happens during the first few second of the experiment, and therefore during this short time the heating process can be considered as adiabatic. Second, because it occurs during the first seconds after the magnetic field is turned on, the maximum slope is located at an absolute temperature close to room temperature, irrespective of the final SPA value. Consequently, all SPA values are estimated at nearly the same (room) temperatures.

The specific heat c of each sample was estimated from the heat capacity of each component in the system (water=4.12, $Fe_3O_4$=0.94, chitosan= 1.19 and PNiPAAm=1.2 J/g*K).

**Figure 6a** shows the results of detailed magnetic field effect on 2 wt% chitosan ferrogels subjected to 30 min inmersion time in NaOH (Series III). As can be observed, the initial temperature rise is fast (i.e., ≈10 ºC/min) and it gradually decreases, although after 20 min of magnetic field applications and at T ≈ 60 ºC, temperature was still increasing. This trend is also observed for Alg-PNiPAAm semi-IPNs (series I and II) as it can be observed in the representative plot corresponding to the sample Alg0.5-PNiPAAm subjected to one oxidation cycle (figure 6b) for which after 20 min of magnetic field application, the temperature reached is T ≈ 45 ºC. It is important to note that this temperature is well above the LCST reported for PNiPAAm in Alg-PNiPAAm semi-IPNs and therefore a rapid deswelling of the samples is observed during the course of the experiment.

Table 2 reports the SPA values (W/$g_{sample}$) obtained for all the samples under study. For each series, the SPA value increases with the iron oxide content since the heat generation is proportional to the amount of magnetic material in each sample. It is worth noting that the SPA values obtained are comparable in all cases to other systems evaluated for magnetic hyperthermia, for example, the SPA reported for a poly (N-isopropylacrylamide) ferrogel (2 wt% $Fe_3O_4$) was 1.45 W/g[30].

The SPA values normalized to the iron oxide content show a different trend as it can be observed in **Figure 7**. For Alg-PNiPAAm semi-IPNs (series I and II), SPA (W/$g_{NPs}$) decreases with the number of oxidation cycles and hence with the iron oxide content being the SPA values corresponding to series II (1 wt% alginate) higher than the corresponding to series I (0.5 wt% alginate). On the contrary, SPA values exhibited by chitosan ferrogels (series III) remain almost constant with the immersion time in alkaline solution being 228 and 211 W/$g_{NPs}$ for the samples subjected to 30 minutes and 60 minutes immersion time in NaOH respectively.

The efficiency as heating agents of iron oxide NPs show a strong dependence of the SPA with the particle size, with maximum absorption of rf power within a definite value $<d>_{max}$. For a given material, the most important parameter determining $<d>_{max}$ is the effective magnetic anisotropy of each material, as inferred from the Néel relaxation model in single domain particles[31]. Additionally, $<d>_{max}$, depends on sample properties like particle agglomeration and polydispersity. For magnetite, the Neel model predicts that $<d>_{max}$ should be located around 25-30 nm [6].

We have previously demonstrated that the average size obtained for the NPs does not vary with the number of oxidation cycles, being around 10 nm for all the samples[20, 21]. Therefore, a feasible explanation for the differences in SPA regarding series I and II is the combined effect of agglomeration and the different magneto crystalline anisotropies of maghemite and magnetite that, for the observed average size of NPs, should give different contributions to the total SPA of the samples. The maghemite/magnetite ratio changes from 1.1 to 1.3 in series II (1 wt% alginate) and from 1.7 to 2.1 in series I (0.5 wt% alginate) when the number of oxidation cycles is increased from 1 to 3 as determined through Confocal Raman Spectroscopy. Hence the results suggest a lower contribution of the maghemite crystalline form (the oxidized form from magnetite) to the total SPA of the system.

## Conclusions

In conclusion, we have demonstrated that the maghemite/magnetite ratio in iron oxide NPs synthesized inside polysaccharide hydrogels through coprecipitation of iron chlorides in an alkaline solution increases with the number of oxidation cycles but it is independent from the immersion time in the alkaline solution. This in turn influences the heating response of the samples to alternating magnetic fields, which points to a lower contribution of the maghemite crystalline form (the oxidized form from magnetite) to the total SPA of the system.

In addition, we have observed high-quality thermal response of the ferrogels synthesized, attaining final temperatures well above the required for the volume phase transition of the Poly(N-isopropylacrylamide) ferrogels, and within short application times (ca 10 min). This makes the present materials an excellent candidate for fast and controlled drug release by external magnetic fields. Experiments with hydrosoluble model drugs for quantifying the net amount of drug release, as well as cytotoxicity effects of these materials, are being performed as the necessary next step for bioapplications of these materials.

**Table I: Composition of the samples**

| SERIES | SAMPLE | $Wt_{H2O}$ (%) | $Wt_{NPs}$ *(%) |
|---|---|---|---|
| I | Alg0.5-PNiPAAm(1c) | 94.9 | 15.7 |
| I | Alg0.5-PNiPAAm(3c) | 91.6 | 30.3 |
| II | Alg1-PNiPAAm(1c) | 92.3 | 7.3 |
| II | Alg1-PNiPAAm(3c) | 90.5 | 17.5 |
| III | Chit2_30min | 95.7 | 13.8 |
| III | Chit2_60min | 95.1 | 14.2 |

**\* Referred to the dried ferrogel**

**Table II**: Magnetic parameters of the samples under study

| SERIES | SAMPLE | $T_B$ | Ms (emu/$g_{Fe3O4}$) | | Hc (Oe) | |
|---|---|---|---|---|---|---|
| | | | 295 K | 5 K | 295 K | 5 K |
| I | Alg0.5-PNiPAAm(1c) | 157(2) | 86 | 90 | 5(5) | 255(5) |
| | Alg0.5-PNiPAAm(3c) | 107(2) | 82 | 90 | 8(5) | 190(5) |
| II | Alg1-PNiPAAm(1c) | 148(2) | 89 | 92 | 7(5) | 230(5) |
| | Alg1-PNiPAAm(3c) | 157(2) | 85 | 87 | 8(5) | 168(5) |
| III | Chit2_30min | 134(3) | 88 | 90 | 4(4) | 195(3) |
| | Chit2_60min | 135(2) | 86 | 89 | 5(5) | 189(3) |

**Table III.** SPA of the samples under study

| SERIES | SAMPLE | wt % $Fe_3O_4$ | SPA (W/g) | SPA (W/g$Fe_3O_4$) |
|---|---|---|---|---|
| I | Alg0.5-PNiPAAm(1c) | 15.7 | 1.78 | 236 |
|   | Alg0.5-PNiPAAm(3c) | 30.3 | 2.39 | 96 |
| II | Alg1-PNiPAAm(1c) | 7.3 | 1.48 | 298 |
|   | Alg1-PNiPAAm(3c) | 17.5 | 1.91 | 119 |
| III | Chit2_30min | 13.8 | 1.35 | 228 |
|   | Chit2_60min | 14.2 | 1.47 | 211 |

**Figure captions**

**Figure 1.** Representative TEM images of iron oxide nanoparticles synthesized in Alg-PNiPAAm semi-IPNs subjected to one oxidation cycle (a) Alg 0.5-PNiPAAm and (b) Alg1-PNiPAAm (Scale bar=200 nm). The inset in both figures shows iron oxide NPs at a higher magnification (Scale bar=10 nm).

**Figure 2.** Comparison of micro-Raman spectra of a) Alg1-PNiPAAm, and Alg1-PNiPAAm subjected to one oxidation cycle (b) and three oxidation cycles (c). Dashed green lines correspond to iron oxide Raman spectra bands fitted with Lorentzian profiles.

**Figure 3.** Comparison of micro-Raman spectra of a) raw Chitosan and chitosan ferrogels (2 wt%) subjected to (b) 30 min immersion time in NaOH and (c) 60 min immersion time in NaOH.

**Figure 4.** Zero-field-cooled (ZFC) and field-cooled (FC) curves for (a) Alg0.5-PNiPAAm (1c) and (b) Alg1-PNiPAAm (3c)

**Figure 5.** Magnetization hysteresis curves measured at (■) 10 K and (○) 295 K for (a) Alg0.5-PNiPAAm (1c) and (b) Alg1-PNiPAAm (1c). Inset: enlargement of the low-field region showing the different coercive fields for the NPa at 10K and 295 K.

**Figure 6.** Temperature increase of (a) chitosan ferrogels (2 wt%) subjected to 30 min immersion time in NaOH and (b) Alg0.5-PNiPAAm subjected to 1 oxidation cycle

**Figure 7.** Specific power absorption of iron oxide NPs synthesized inside Alg-PNiPAAm ferrogels (series I and II) and chitosan ferrogels (series III).

**FIGURE 1**

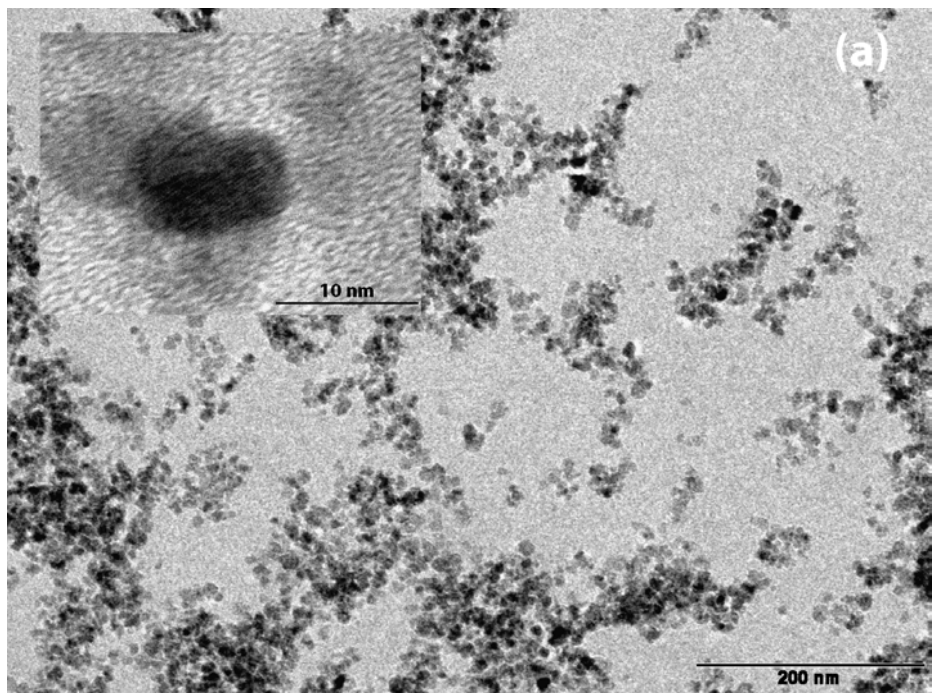

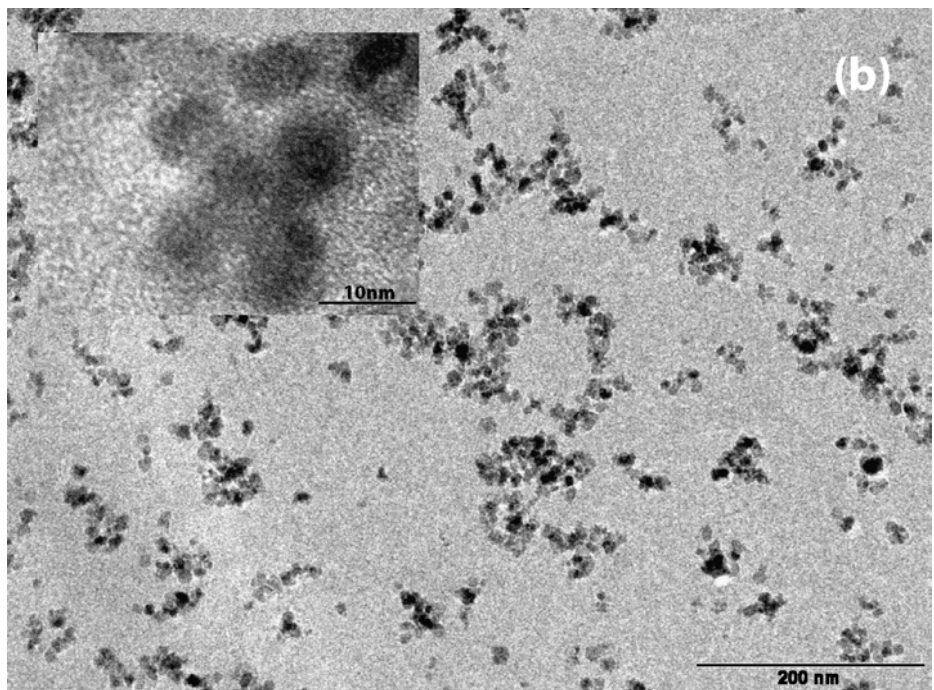

**FIGURE 2**

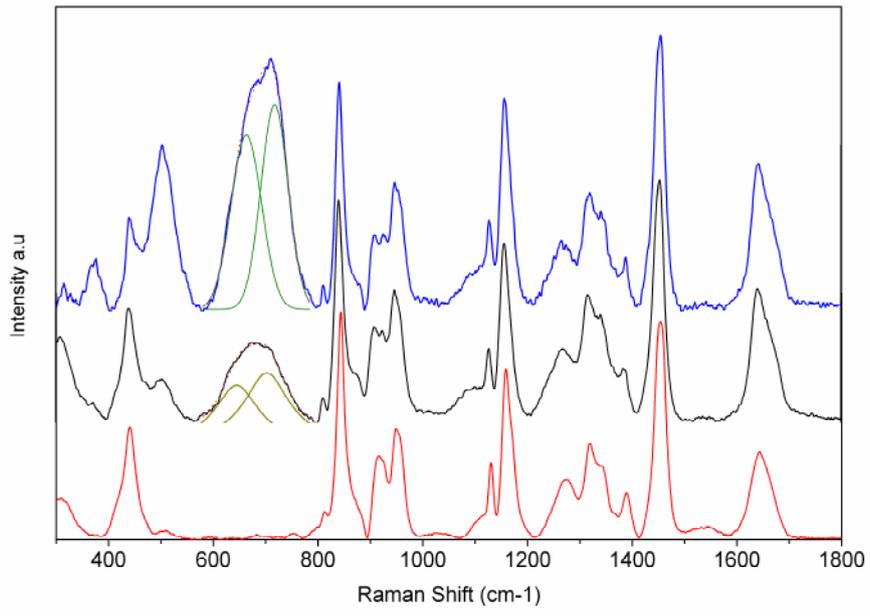

**FIGURE 3**

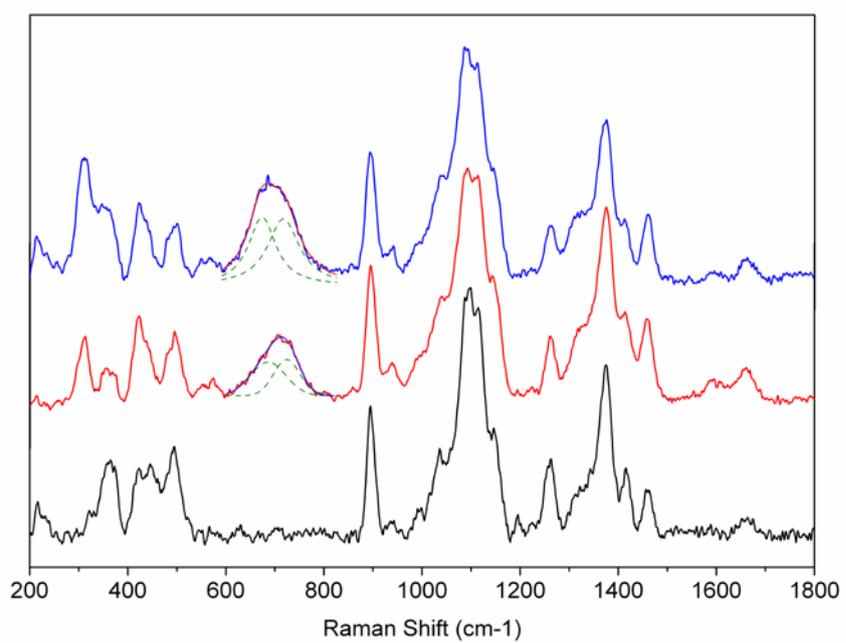

**FIGURE 4.**

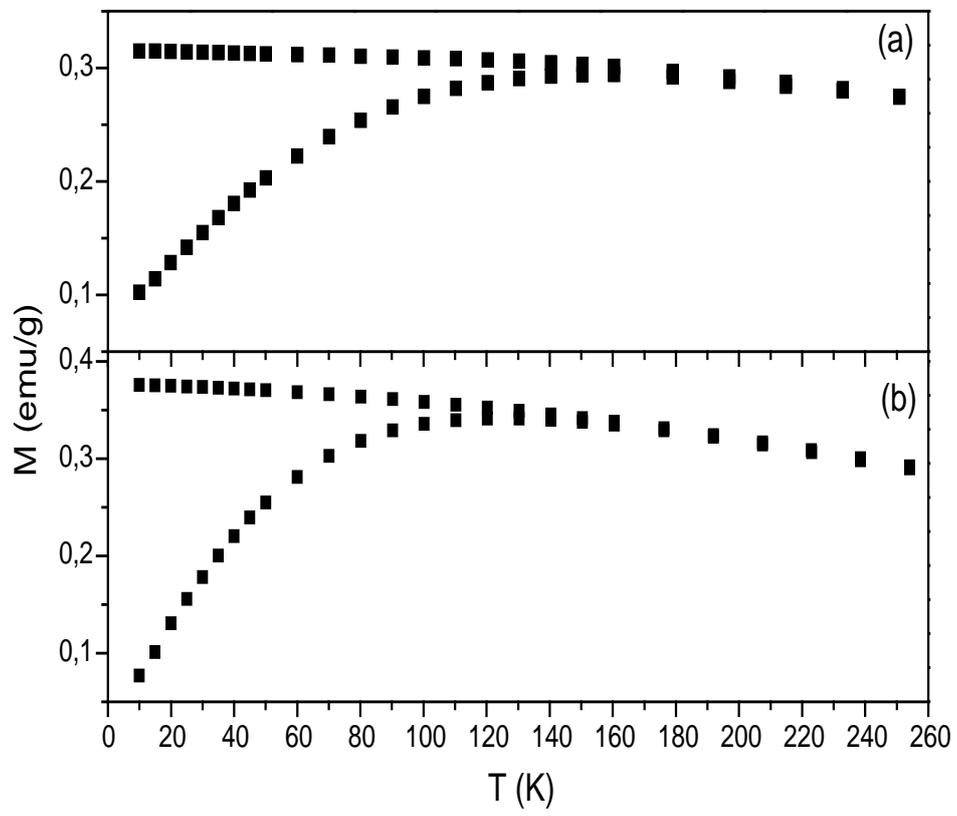

**FIGURE 5.**

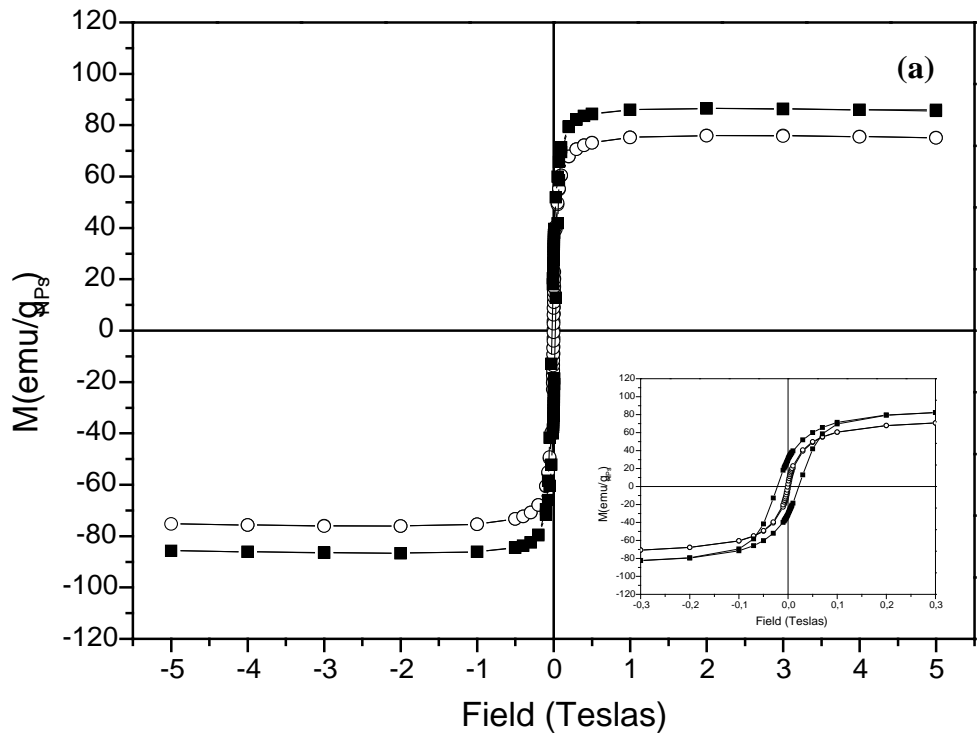

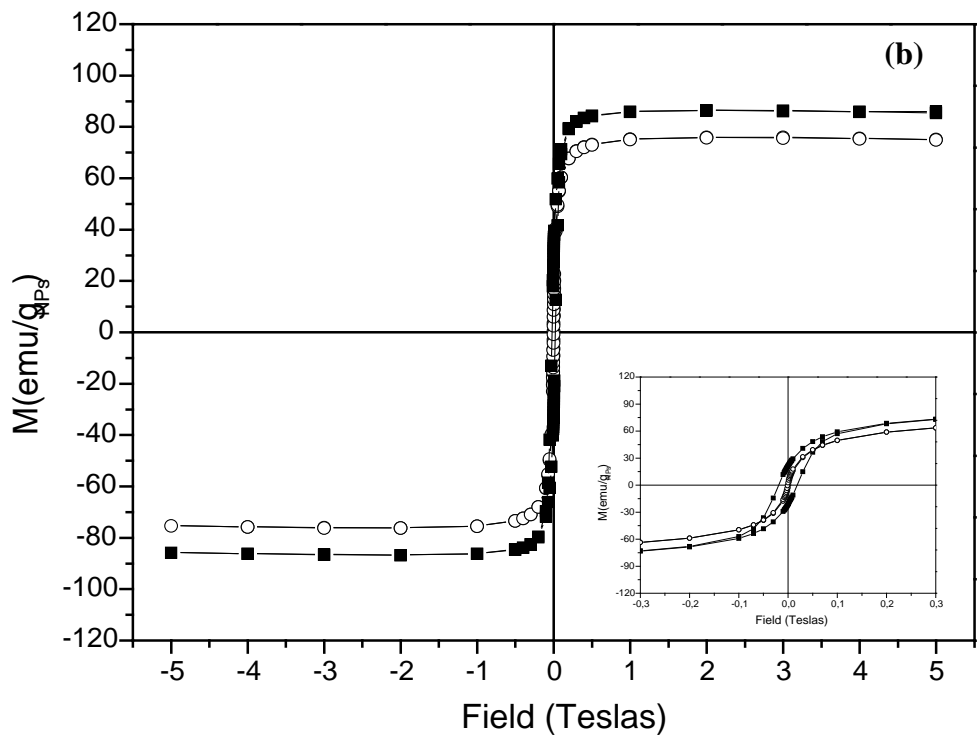

**FIGURE 6**

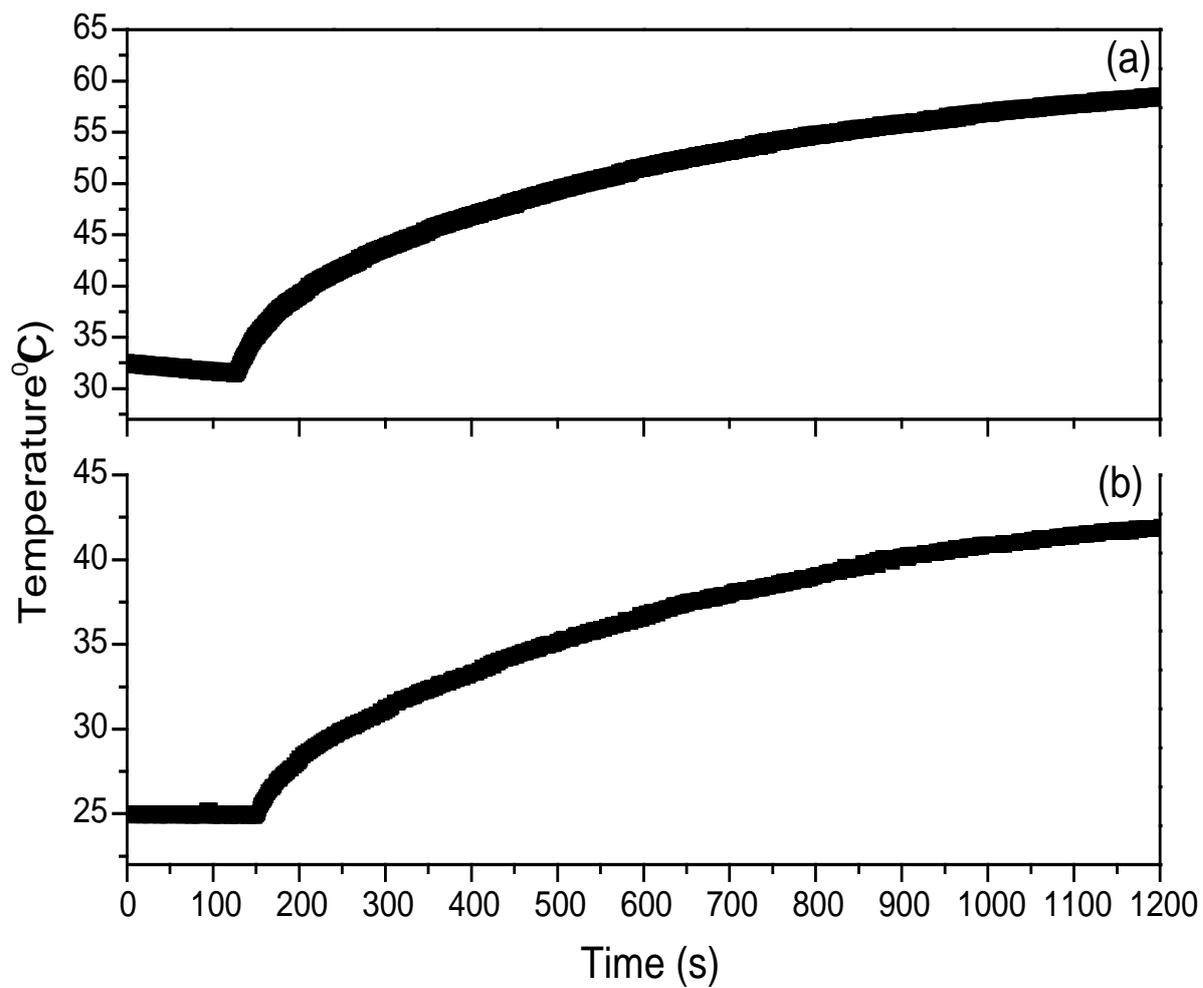

**FIGURE 7**

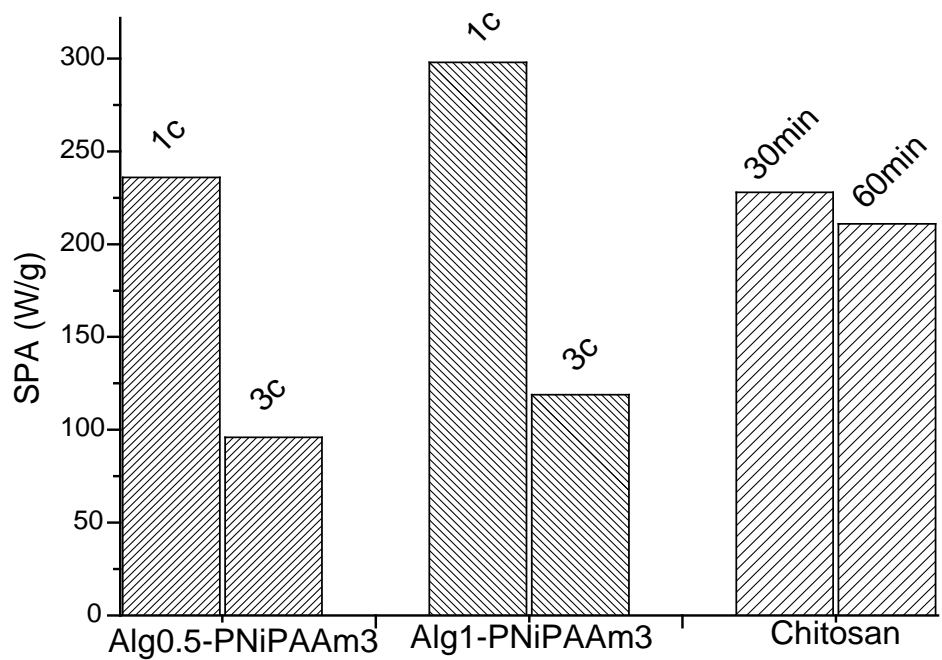